\documentclass[twocolumn,preprintnumbers,amsmath,amssymb]{revtex4-1}
\usepackage{epsfig}
\usepackage{amsmath}
\usepackage{array}
\usepackage{multirow}
\usepackage{psfrag}
\usepackage[font=footnotesize,format=plain,labelsep=period,justification=raggedright]{caption}
\usepackage{subcaption}
\usepackage{graphicx}

\begin{document}
\title{A study of the ice-water interface using the TIP4P/2005 water model}
\author{Jorge Benet, Luis G. MacDowell and Eduardo Sanz}
\affiliation{Departamento de Qu\'{\i}mica F\'{\i}sica,
Facultad de Ciencias Qu\'{\i}micas, Universidad Complutense de Madrid,
28040 Madrid, Spain}
\date{\today}

\begin{abstract}

In this work we study the ice-water interface under coexistence conditions by means of molecular simulations
using the TIP4P/2005 water model.  Following the methodology proposed by Hoyt
and co-workers [J. J. Hoyt, M. Asta and A. Karma, Phys. Rev. Lett., {\bf 86},
5530, (2001)] we measure the interfacial free energy of ice with liquid water
by analysing the spectrum of capillary fluctuations of the interface.  We get
an orientationally averaged interfacial free energy of 27(2) mN/m, in good 
agreement with a recent estimate obtained from simulation data of the size of 
critical clusters [E. Sanz, C. Vega, J. R. Espinosa, R. Caballero-Bernal, J. L. F. Abascal and C. Valeriani, JACS, {\bf 135}, 15008, (2013)]. We also 
estimate the interfacial free energy of different planes and obtain 27(2), 28(2) 
and 28(2) mN/m for the basal, the primary prismatic and the secondary prismatic planes respectively. 
Finally, we inspect the structure of the interface and find that its thickness 
is of approximately 4-5 molecular diameters. Moreover, we find that when 
the basal plane is exposed to the fluid the interface alternates regions of cubic ice with 
regions of hexagonal ice.

\end{abstract}

\maketitle
\renewcommand{\arraystretch}{1.5}

\section{Introduction}
The interfacial free energy between ice and water, $\gamma_{iw}$, is a crucial parameter in  
ice nucleation and growth \cite{pruppacher1995,pruppacher1967}. 
Despite its importance, there is not yet a well established experimental value 
for $\gamma_{iw}$. The spread of experimental data for $\gamma_{iw}$, ranging from 25 to 35 mN/m \cite{pruppacher1995},  
sharply contrasts with the accuracy with which the interfacial free energy of the liquid-vapour 
interface is known \cite{expsurfacetensionwater}. Unfortunately, there is no accurate experimental technique for the
determination of the crystal-melt interfacial free energy. 

In order to aid experimentalists in finding a definite value for $\gamma_{iw}$, 
guidance from computer simulation is highly valuable. 
However, there are not many studies devoted to the estimation of 
$\gamma_{iw}$ from simulations. 
Recently, 
$\gamma_{iw}$ has been calculated for a series of water models with \cite{doi:10.1021/ct300193e} and
without \cite{PhysRevLett.100.036104} taking full electrostatic interactions into account. 
In these works, a variant of the cleaving method \cite{broughton:5759} was used to compute $\gamma_{iw}$ 
and the studied models were TIP4P, TIP4P-Ew and TIP5P-E. 

There are numerous water models currently available in the literature with which 
different predictions of the behaviour of real water can be made. 
In a recent work, Abascal and Vega have compared the ability of many different
rigid, non-polarizable models to predict a comprehensive set of real water properties. 
The TIP4P/2005 model \cite{JCP_2005_123_234505} turned out to be the one that does the best job in the 
overall description of the behaviour of real water \cite{Vega11}. 
Therefore, estimating $\gamma_{iw}$ for such model would be highly relevant.  

In a recent publication by some of the authors of this work, 
$\gamma_{iw}$ was estimated for the TIP4P/2005 model \cite{jacs2013} with a 'seeding' method
originally used by Bai and Li 
to study the crystal-melt interface of the the Lennard-Jones system \cite{bai:124707}.  This method consists in measuring the critical size of crystalline
clusters and then obtaining $\gamma_{iw}$ from Classical Nucleation Theory \cite{ZPC_1926_119_277_nolotengo,becker-doring}. 
Therefore, this method provides an indirect estimate of $\gamma_{iw}$. 
Moreover, the method by Bai and Li does not provide information about the dependency of $\gamma_{iw}$
with the orientation of the crystal, since an orientationally averaged 
$\gamma_{iw}$ is obtained. 

In this paper we evaluate $\gamma_{iw}$ for the TIP4P/2005 model by means of the Capillary Fluctuation Method \cite{PhysRevLett.86.5530}.
This method has been used, for instance, for the calculation of the
interfacial free energy of hard spheres \cite{JCP_2006_125_094710}, the 
Lennard-Jones \cite{morris:3920} and dipolar fluids \cite{wangmorrisJCP2013}.
Here we evaluate $\gamma_{iw}$ for 
the TIP4P/2005 model for the basal, prismatic I and prismatic II planes of ice. 
We find an average value of $\gamma_{iw}$ of 27(2) mN/m and a small anisotropy 
between different orientations. 
Finally, we inspect the structure of the interface. We estimate the thickness of the 
interface to be of about 4-5 molecular diameters. Moreover, we find that when the basal 
plane is exposed to the liquid the interface develops alternating hexagonal and cubic ice regions. 

\begin{figure}
 \includegraphics[width=0.25\paperwidth,height=0.33\paperwidth,keepaspectratio]{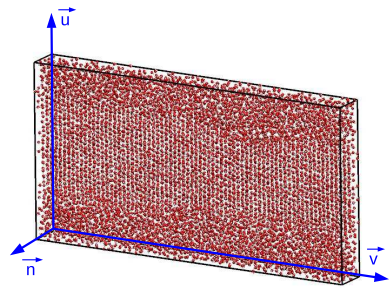}
 \caption{Snapshot of a typical configuration. Only oxygen atoms are shown. Particles are coloured in orange if they have a solid-like
local environment and in blue otherwise. The edges of the simulation box and the vectors that define the orientation of the crystal and the propagation direction of capillary
waves are shown in the figure.}
 \label{caja}
\end{figure}

\section{Methods}

We use the Capillary Fluctuation Method \cite{PhysRevLett.86.5530} to 
compute the interfacial free energy. The method consists in measuring 
$\gamma_{iw}$ by analysing the profile of the 
interface between ice and water under coexistence conditions. For the TIP4P/2005
model the interface between ice and water is rough, as can be seen in Fig. 
\ref{caja}, where particles in the ice phase are shown in orange and 
particles in the liquid phase are shown in blue. Particles are labelled as
ice or liquid-like based on local bond order parameters \cite{lechnerDellago08,benetjcp2014}. 
By knowing which particles
belong to each phase, one can establish a discretized interface profile
along the $x$ direction, $h(x_n)$ (in Ref. \cite{benetjcp2014} a detailed explanation 
of the way we establish $h(x_n)$ is given). Then, $h(x_n)$ is Fourier-transformed, 
\begin{equation}
 h_q=\frac{1}{N}\overset{N}{\underset{n=1}{\sum}} h(x_n)e^{iqx_n}, 
 \label{Fourier_transform}
\end{equation}
and an amplitude, $h_q$, is obtained for each wave vector, $q$, where 
$q$ is a multiple of $2\pi/L_x$. 
Small $q$ vectors correspond to wave modes with a large wave length and vice-versa.
In the equation above $N$ is
the number of discretization points along the $L_x$ side of the simulation box. 

Through the equipartition theorem, Capillary Wave Theory provides the following
relation between $h_q$ and the interfacial stiffness, $\widetilde \gamma$
\cite{fisher83,privman92,jasnow84,nelson04}:
\begin{equation}
 \left<|h_q|^2\right>=\frac{k_BT}{A\widetilde \gamma q^2}
\label{eqstiffness}
\end{equation}
where $A=L_x \cdot L_y$ is the interfacial area, (see Fig. \ref{caja}). 
The calculated stiffness depends on the crystal plane that is exposed to the fluid
and on the direction along which the wave propagates. 
The exposed crystal plane is perpendicular to the vector $\vec{u}$ in Fig. \ref{caja}
and it is identified by its Miller indices. 
The direction of propagation of the wave is perpendicular to both $\vec{u}$ and $\vec{n}$
and it is specified by the Miller indices of the plane perpendicular to $\vec{n}$. 
Hence, $\widetilde \gamma \equiv \widetilde \gamma(\vec{u},\vec{n})$. 

\begin{figure}
\includegraphics[width=0.4\paperwidth,keepaspectratio]{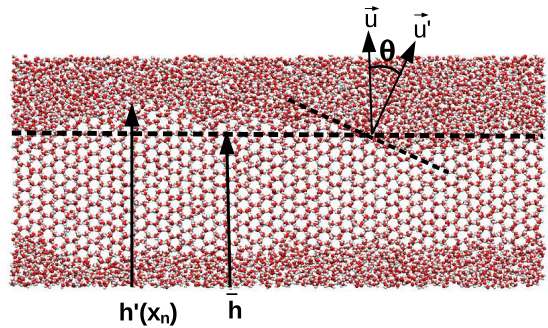}
\caption{Snapshot of a configuration of an ice slab in equilibrium with liquid
water. $h(x_n)$ is the interfacial height, 
$\vec{u}$ is the vector perpendicular to the average interface position, $\vec{u'}$ is 
the vector perpendicular to the instantaneous interface and $\theta$ is the 
angle between $\vec{u}$ and $\vec{u'}$.}
\label{hx}
\end{figure}

Once the stiffness is known, we use the relation \cite{fisher83}:
\begin{equation}
\widetilde{\gamma}(\vec{u},\vec{n})=\left ( \gamma(\theta) + \frac{d^2 \gamma 
(\theta)}{d \theta ^2} \right )_{\theta =0}
\label{gammastiff}
\end{equation}
to obtain the interfacial free energy. 
In the above expression $\theta$ is the angle between the average planar interface defined by
$\vec{u}$  and the vector normal to the instantaneous interface $\vec{u'}$. The definition of
$\theta$ is sketched in Fig. \ref{hx}. 

\begin{figure}
\includegraphics[width=0.33\paperwidth,keepaspectratio]{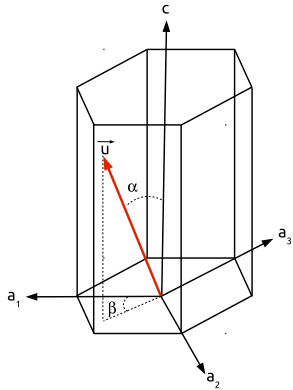}
\caption{Hexagonal reference system for the ice Ih structure.
The vector $\vec{u}$ determines the crystallographic plane exposed at the interface and it is 
characterized by the angles $\alpha$ and $\beta$.}
\label{sistema_hexagonal}
\end{figure}

Obtaining $\gamma$ from Eq. \ref{gammastiff} requires 
first defining the dependence of the interfacial free energy with the orientation of the crystal, $\gamma(\vec{u})$. 
Since the point group of hexagonal ice is 6/$mmm$, the orientation dependence of 
$\gamma(\vec{u})$ can be written as an expansion in terms of Spherical Harmonics \cite{kara81}:
\begin{equation}
\begin{aligned}
\gamma(\vec{u})/\gamma_0 \approx &1+\epsilon_1y_{20}(\alpha,\beta)+\epsilon_2y_{40}(\alpha,\beta)\\
&+\epsilon_3y_{60}(\alpha,\beta)+\epsilon_4y_{66}(\alpha,\beta)+...
\end{aligned}
\label{gamma_hexagonal}
\end{equation}
where $\gamma_0$ is the interfacial free energy averaged over all orientations, $\alpha$ and 
$\beta$ are the spherical angles defining a given plane (see Fig. 
\ref{sistema_hexagonal}) and $\epsilon _k$ are the anisotropy parameters. 
The functions $y_{lm}(\alpha, \beta)$ are the normalized 
spherical harmonics, and they are provided in Table \ref{spherical_harmonics}.
In Table \ref{expresiones_gamma_hexagonal}, Eq. \ref{gamma_hexagonal} is expressed      
for the particular case of the three orientations of ice-Ih we put in contact with liquid
water in this work. 

\begin{table}
\centering
\scalebox{0.90}{\begin{tabular}{l}
\hline
\hline
$y_{20}(\alpha,\beta)=\sqrt{5/16\pi}\left[ 3cos^2(\alpha)-1 \right]$\\
$y_{40}(\alpha,\beta)=\frac{3}{16}\sqrt{1/\pi}\left[ 35cos^4(\alpha)-30cos^2(\alpha)+3 \right]$\\
$y_{60}(\alpha,\beta)=\frac{1}{32}\sqrt{13/\pi}\left[ 231cos^6(\alpha)-315cos^4(\alpha)+105cos^2(\alpha)-5 \right]$\\
$y_{66}(\alpha,\beta)=\frac{1}{64}\sqrt{6006/\pi}\left[ 1-cos^2(\alpha)\right]^3cos(6\beta)$\\
\hline
\hline
\end{tabular}}
\caption{Expressions for the normalized spherical harmonics used in
Eq.\ref{gamma_hexagonal}.}
\label{spherical_harmonics}
\end{table}

\begin{table}
\centering
\scalebox{0.80}{\begin{tabular}{c c}
\hline
\hline
Interfacial plane &  $\gamma(\vec{u})/\gamma_0$ \\
\hline
\\
(0001)   & $1+\frac{1}{2}\sqrt{5/\pi}\epsilon_{1}+\frac{3}{2}\sqrt{1/\pi}\epsilon_{2}+\frac{1}{2}\sqrt{13/\pi}\epsilon_{3} $\\
(10\={1}0) & $1-\frac{1}{4}\sqrt{5/\pi}\epsilon_{1}+\frac{9}{16}\sqrt{1/\pi}\epsilon_{2}-\frac{5}{32}\sqrt{13/\pi}\epsilon_{3}-\frac{1}{64}\sqrt{6006/\pi}\epsilon_{4} $\\
(11\={2}0) & $1-\frac{1}{4}\sqrt{5/\pi}\epsilon_{1}+\frac{9}{16}\sqrt{1/\pi}\epsilon_{2}-\frac{5}{32}\sqrt{13/\pi}\epsilon_{3}+\frac{1}{64}\sqrt{6006/\pi}\epsilon_{4} $\\
\hline
\hline
\end{tabular}}
\caption{Interfacial free energy expansion in terms of spherical harmonics for the different crystallographic planes studied in this work. 
(0001) corresponds to the basal plane, (10\={1}0) to the primary prismatic and
(11\={2}0) to the secondary prismatic.}
\label{expresiones_gamma_hexagonal}
\end{table}

By taking the second derivative of Eq. \ref{gamma_hexagonal} with respect to $\theta$ and plugging the result into 
Eq. \ref{gammastiff} an expansion of $\widetilde{\gamma}$ is obtained. 
Such expansion is given in Table \ref{expansion_stiff} for all the orientations studied
in this work. The equations in Table \ref{expansion_stiff} combined with the simulation 
results for $\widetilde{\gamma}$ allow for the calculation of
$\epsilon_k$ and $\gamma_0$. With these, the interfacial free energy is obtained 
with the equations provided in Table \ref{expresiones_gamma_hexagonal}.

In summary, we simulate the interface under coexistence conditions and obtain 
an average amplitude, $h_q$, for each wave vector, $q$, via Eq. \ref{Fourier_transform} by defining 
an interfacial profile, $h(x_n)$, for many independent configurations. 
Then, $\widetilde \gamma (q)$ is obtained by means of Eq. \ref{eqstiffness}. 
Once $\widetilde \gamma$ has been calculated for different orientations we
solve the system of equations given in Table \ref{expansion_stiff} to obtain 
$\gamma_0$ and the anisotropy parameters $\epsilon_k$. Finally, we use the calculated
$\gamma_0$ and $\epsilon_k$ to obtain the interfacial free energy of each plane via the expressions given in 
Table \ref{expresiones_gamma_hexagonal}.

\begin{table}
\centering
\scalebox{0.87}{\begin{tabular}{c c c}
\hline
\hline
$\vec{u}$ & $\vec{n}$ & $\widetilde{\gamma}(\vec{u},\vec{n})/\gamma_0$ \\
\hline
0001    & 11\={2}0     & $1-\sqrt{5/\pi}\epsilon_{1}-\frac{27}{2}\sqrt{1/\pi}\epsilon_{2}-10\sqrt{13/\pi}\epsilon_{3} $\\
10\={1}0& 11\={2}0         & $1+\frac{5}{4}\sqrt{5/\pi}\epsilon_{1}-\frac{171}{16}\sqrt{1/\pi}\epsilon_{2}+\frac{205}{32}\sqrt{13/\pi}\epsilon_{3}+\frac{5}{64}\sqrt{6006/\pi}\epsilon_{4} $\\
10\={1}0& 0001 & $1-\frac{1}{4}\sqrt{5/\pi}\epsilon_{1}+\frac{9}{16}\sqrt{1/\pi}\epsilon_{2}-\frac{5}{32}\sqrt{13/\pi}\epsilon_{3}+\frac{35}{64}\sqrt{6006/\pi}\epsilon_{4} $\\
11\={2}0& 10\={1}0   & $1+\frac{5}{4}\sqrt{5/\pi}\epsilon_{1}-\frac{171}{16}\sqrt{1/\pi}\epsilon_{2}+\frac{205}{32}\sqrt{13/\pi}\epsilon_{3}-\frac{5}{64}\sqrt{6006/\pi}\epsilon_{4} $\\
11\={2}0& 0001 & $1-\frac{1}{4}\sqrt{5/\pi}\epsilon_{1}+\frac{9}{16}\sqrt{1/\pi}\epsilon_{2}-\frac{5}{32}\sqrt{13/\pi}\epsilon_{3}-\frac{35}{64}\sqrt{6006/\pi}\epsilon_{4} $\\
\hline
\hline
\end{tabular}}
\caption{Expansions for the stiffness for the different orientations studied.}
\label{expansion_stiff}
\end{table}

\subsection{Simulation details}
To simulate our system we have employed the 
Molecular Dynamics package GROMACS \cite{GROMACS1, GROMACS2}. Production runs for a total 
time of $\sim0.5 \mu s$ were carried out in the NVT ensemble with the 
time step for the Velocity-Verlet integrator fixed to 0.003 ps, and 
snapshots were saved every 75 ps. The temperature was set to 248.5 K (close to 
the reported melting temperature of the model \cite{JCP_2005_123_234505}) and the density was fixed
close to an average value between the coexistence densities at 1 bar of liquid water and ice-Ih. 
At these thermodynamic conditions
the interface fluctuates but the relative ice/water amount stays constant throughout the simulation. 
To fix the temperature we employed a 
velocity-rescale thermostat \cite{bussi07} with a relaxation time of 2 ps.

\begin{table}[h]
\footnotesize
\begin{tabular}{c c c c c}
\hline
\hline
Orientation & $L_x$x$L_y$x$L_z (nm^3)$ & Molecules\\
\hline
(Basal)[pII] & 18.7696x1.8039x9.3319 & 10112 \\
(pI)[Basal]  & 18.0134x2.1991x8.0808 & 10240 \\
(pI)[pII]    & 17.6430x2.3491x7.8227 & 10368 \\
(pII)[Basal] & 17.9927x2.2047x8.3875 & 10670 \\
(pII)[pI]    & 18.3690x1.8037x8.3928 & 8896  \\
\hline
\hline
\end{tabular}
\caption{System size for all ice-water orientations studied in this work.}
\label{sizes-ice-water}
\end{table}

An initial configuration in which water and ice coexist at 1 bar is prepared as 
described in Ref. \cite{benetjcp2014}. 
The $L_x$ and $L_y$ axis of the simulation box are carefully chosen to avoid
any stress in the crystal lattice \cite{benetjcp2014,Frenkel_darkside}.
The size and crystal orientation of the simulated systems are summarized in 
Table \ref{sizes-ice-water}. 
The box geometry with a long $x$ axis (see Fig. \ref{caja}) allows for the study of long wave-length capillary
waves without having a prohibitively large number of molecules in the system. 
Moreover, it allows 
to easily control the direction of wave propagation. 
It has been shown that the chosen box geometry with a large $L_x/L_y$ ratio gives
the same stiffness as boxes with $L_x/L_y$ close to 1 \cite{0295-5075-93-2-26006,benetjcp2014}.

\section{Results}

\subsection{Stiffness}

\begin{figure}
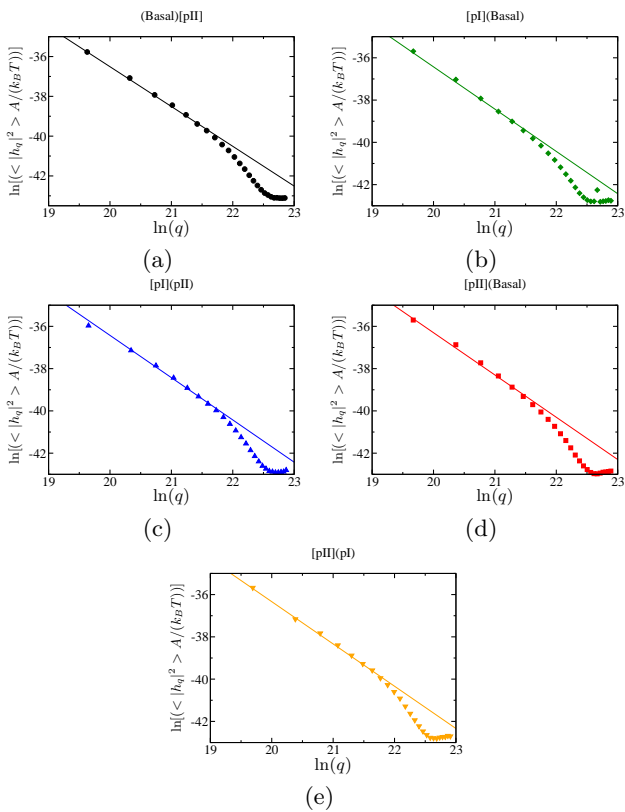

\psfrag{lnq}[][Bc][5]{{$\ln(q)$}}
\psfrag{lnampli}[][tc][4]{{$\ln[(<|h_q|^2>A/(k_BT))$]}}
\centering
\begin{subfigure}[b]{0.15\paperheight}
\resizebox{1.5in}{!}{\includegraphics{fig4.eps}}
\caption{}
\end{subfigure}
\begin{subfigure}[b]{0.15\paperheight}
\resizebox{1.5in}{!}{\includegraphics{fig5.eps}}
\caption{}
\end{subfigure}
\begin{subfigure}[b]{0.15\paperheight}
\resizebox{1.5in}{!}{\includegraphics{fig6.eps}}
\caption{}
\end{subfigure}
\begin{subfigure}[b]{0.15\paperheight}
\resizebox{1.5in}{!}{\includegraphics{fig7.eps}}
\caption{}
\end{subfigure}
\begin{subfigure}[b]{0.15\paperheight}
\resizebox{1.5in}{!}{\includegraphics{fig8.eps}}
\caption{}
\end{subfigure}
 \caption{Plots of  $\ln[<|h_q|^2>A/(k_BT)]$ vs $\ln(q)$ for all ice-water orientations
studied in this work. $<|h_q|^2>A/(k_BT)$ is given in m$^3$/N and $q$ is given in m$^{-1}$. Symbols are our 
simulation data and straight lines are linear fits of slope minus 2 to the
low-$q$ data. The intercept of such fits is $-\ln(\widetilde \gamma)$.}
 \label{ajustes-sl}
\end{figure}

By simulating the interface for a long time ($\sim$ 0.5 $\mu$s) we gather 
thousands of configurations and obtain interfacial profiles, $h(x_n)$,
for each of the two ice-water interfaces present in the simulation
box.  Then we Fourier-transform each $h(x_n)$ (Eq.
\ref{Fourier_transform}) to obtain estimates of $|h_q|^2$,
which we average to get $<|h_q|^2>$. 
For a rough interface, by representing $\ln[<|h_q|^2>A/(k_BT)]$ vs $\ln(q)$ we
should obtain, in the $q$ regime where Eq. \ref{eqstiffness} holds, a straight
line of slope minus 2 and intercept $-\ln(\widetilde \gamma)$. Such plots are shown
in Fig. \ref{ajustes-sl} for all orientations studied in this work. Symbols
correspond to our simulation data and straight lines to a linear fit with slope
minus 2 to the low-$q$ points. As expected from Eq. \ref{eqstiffness} the fit
describes quite well our data, at least for the low-$q$ regime, allowing us to get $\widetilde
\gamma$ from the intercept.  The values of $\widetilde \gamma$ thus obtained
are reported in Table \ref{stiffness-sl}. 

\begin{table}
\footnotesize
\begin{tabular}{c c}
\hline
\hline
Crystal Orientation & $\widetilde{\gamma}_{iw}(mN/m)$ \\
\hline
(Basal)[pII] & 29.8 \\
(pI)[Basal]  & 28.1 \\
(pI)[pII]    & 28.1 \\
(pII)[Basal] & 24.7 \\
(pII)[pI]    & 25.1 \\
\hline
\hline
\end{tabular}
\caption{Stiffness of all ice-water orientations studied in this work.}
\label{stiffness-sl}
\end{table}

Notice that the very good fit of the capillary wave spectrum to
Eq. \ref{eqstiffness} indicates that all three crystal faces
studied are rough.  This observation is further confirmed by 
visual inspection of snapshots, as can be seen in Figs. \ref{caja} and 
\ref{hx}. 
Experimental studies on the other hand
indicate that ice crystals in coexistence with water at about the triple 
point  have a faceted basal plane, and a completely circular perimeter
\cite{maruyama97}. Such observation is compatible with prismatic planes which 
are rough, but indicate a basal plane that is below the roughening transition
even at the triple point. Even though the experiments of Ref. \cite{maruyama97} suggest
that the basal plane is not rough, at least for the lengthscales accessible to our simulations
the basal plane shows a rough character that enables the calculation of 
its stiffness and its interfacial free energy by means of the Capillary Fluctuation Method.

\subsection{Interfacial free energy}

Once the stiffness is known for a set of different orientations, we can obtain
the interfacial free energy by solving the system of equations given in Table
\ref{expansion_stiff} and working out the anisotropy parameters, $\epsilon_k$,
and the orientationally averaged $\gamma$, $\gamma_0$. With $\epsilon_k$ and
$\gamma_0$ one can obtain the interfacial free energy for each crystal plane
via the expressions given in Table \ref{expresiones_gamma_hexagonal}. 
Unfortunately, the equations
of Table III are not linearly independent, and it is not 
possible to obtain all 4 anisotropy parameters plus $\gamma_0$. 
In Ref. \cite{sun06} Sun {\it et al.} dealt with a similar problem
in their study of the crystal/melt interface of Mg, which also exhibits
a crystal structure with hexagonal point group symmetry. In this study,
it was noticed that some of the $\epsilon _k$ hardly contributed to
the anisotropy, and could be safely set equal to 0, such that
the  stiffness data could be accurately fitted with the remaining 
$\epsilon _k$. Specifically, it was shown 
that $\epsilon _1$ was necessary to obtain an accurate fit
and that the anisotropy parameter $\epsilon_4$ was necessary to resolve the
anisotropy. 
The other two anisotropy parameters, $\epsilon _2$ and $\epsilon _3$, 
were made equal to zero. Despite the rather different substance studied,
our data are completely consistent with this observation, and we
have therefore followed the same approach. With this strategy, 
we obtain an orientationally averaged interfacial free energy for the 
TIP4P/2005 model of $\gamma_0=27(2)$mN/m. This is in good agreement with the
value of $\gamma_0=29(3)$mN/m recently estimated from measurements of the 
critical nucleus size for
the same model \cite{jacs2013}.  It is also similar to the value of $\gamma_0$
obtained for other water models in Ref. \cite{doi:10.1021/ct300193e}. 
In fact, an average of the $\gamma_{iw}$ calculated for different planes in Ref. \cite{doi:10.1021/ct300193e} 
gives 26.5 mN/m for the TIP4P model and 27.5 mN/m for the TIP4P-Ew.   The
comparison with the experiment is not so straightforward as there is not a
definite experimental value for $\gamma_0$. There are published values ranging
from 25 to 35 mN/m \cite{pruppacher1995,hardy77}. The only thing we can say is
that the value we get for the TIP4P/2005 model is at least within the range of
the reported experimental values.  We have also calculated the interfacial free
energy of the different planes and show the results in Table \ref{gamma-sl}.
We observe a small anisotropy between different planes. It seems that the basal
plane has the smallest interfacial free energy.  However, the uncertainty of
our calculations does not allow us concluding anything definite in this
respect.  In Table \ref{gamma-sl} we also compare our results with those 
obtained in Ref. \cite{doi:10.1021/ct300193e} for the TIP4P and TIP4P-Ew models. 
The similarity between all TIP4P family models is quite strong and, within
the error bar, all models give the same interfacial free energy.

\begin{table}
\footnotesize
\begin{tabular}{c c c c}
\hline
\hline
Crystal Orientation & TIP4P/2005 & TIP4P    & TIP4P-Ew\\
\hline
Basal               &    27(2)            & 24.5(6)  & 25.5(7) \\
Prismatic I         &    28(2)            & 27.6(7)  & 28.9(8) \\
Prismatic II        &    28(2)            & 27.5(7)  & 28.3(7) \\
\hline
\hline
\end{tabular}
\caption{Interfacial free energy of the ice-water interface, in mN/m, for 
different crystal orientations and water models. Values for the TIP4P and TIP4P-Ew have been taken from 
Ref. \cite{doi:10.1021/ct300193e}.}
\label{gamma-sl}
\end{table}

\subsection{Interface structure}

\subsubsection{Density profile}

In order to analyze the structure of the interface
we measure the density profile along the $z$ direction, 
perpendicular to the interface. Such a study must be
taken with some caution, however. The width of the
interface has an intrinsic contribution, that is
characteristic of the substance studied, but also shows
an additional capillary wave term, that depends logarithmically
on the interface area.\cite{weeks77,ocko94,werner97,macdowell14} 
For that reason, average profiles
extracted from a simulation are not strictly intrinsic properties
of the substance, but also depend on the system dimensions.
Since the capillary roughening shows a logarithmic dependence
on the lateral dimensions, however, the correction to the
intrinsic contribution that is typical in a finite simulation box is
quite small. Be as it may, the results that are obtained set an
upper bound for the intrinsic contribution. Furthermore, since
all faces studied have a rather similar lateral dimension, 
the comparison between different crystal orientations also
remains meaningful despite the capillary wave roughening.

\begin{figure}
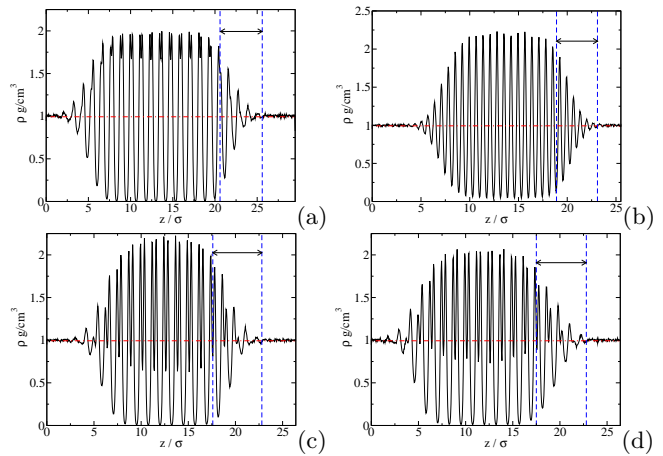

\centering
\resizebox{1.5in}{!}{\includegraphics{fig9.eps}}(a)
\resizebox{1.5in}{!}{\includegraphics{fig10.eps}}(b)
\resizebox{1.5in}{!}{\includegraphics{fig11.eps}}(c)
\resizebox{1.5in}{!}{\includegraphics{fig12.eps}}(d)
\caption{Density profile along the $z$ direction (perpendicular to the interface) for four different orientations:            
(a) (Basal)[pII]; (b) (pII)[basal]; 
(c) (pI)[basal]; and (d) (pI)[pII]. 
We calculate density profiles with slabs of thickness 0.05$\sigma$. Horizontal dotted-dashed lines correspond to the average bulk density of the fluid phase. 
Vertical dashed lines correspond to the approximate location of the interface borders.}
\label{densprof}
\end{figure}

In Fig. \ref{densprof} the density is plotted
along the $z$ direction for four different orientations
corresponding to the basal plane and to both prismatic planes. As a consequence of the 
geometry of our simulations (see Fig. \ref{caja}) two interfaces can be observed for each system. 
To obtain these density profiles we use bins of 0.05 $\sigma$ and average over a time gap 
of $\sim$ 35 ns. 
Using a small bin width allows us to observe the
different crystal layers along the system. 
The horizontal dotted-dashed lines in Fig. \ref{densprof} correspond to the average bulk density of
the fluid phase. As it should be, the density 
given by the profile coincides with the bulk density in the middle of the phase. 
Profiles given in Figs. \ref{densprof} (c) and (d) correspond 
to two different wave propagation directions for the same interfacial plane (the primary 
prismatic plane). As expected, both density profiles are equivalent. 
Note that the profile corresponding to the basal plane (Fig. \ref{densprof} (a)) shows the twin peaks characteristic
of hexagonal planes. 

The measure of the thickness of the interface is a somewhat arbitrary 
task since one has to establish a criterion to locate the interface borders. 
In order to determine these borders we
consider that the interface begins when a density peak does not reach 
the 90\% of the average peak height in the middle of the 
crystal slab, and that it ends  
when the density profile becomes flat. We show the interfacial borders thus
obtained as dashed vertical lines in Fig. \ref{densprof}. We obtain 
an interfacial width of about 4-5 molecular diameters for all studied planes. 
These values are similar, but somewhat larger than the $\sim$ 3 molecular diameters
reported in Ref. \cite{haymetjcp1988} for another TIP4P family model (the TIP4P) and a different system size.

\subsubsection{Hexagonal versus cubic ice}

\begin{figure}
\includegraphics[width=0.25\paperwidth,keepaspectratio,clip]{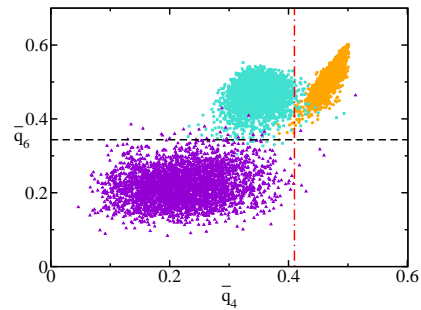}
\caption{Values of the $\bar{q}_{6}$ versus the $\bar{q}_{4}$ 
order parameter \cite{lechnerDellago08} for 3600 molecules of the bulk liquid (magenta), ice Ih (cyan), and ice Ic (orange) phases 
equilibrated under coexistence conditions (1 bar and 250 K). A cut-off distance of 3.5 \AA was used to calculate the order parameter. 
The dashed black line represents  the threshold used to distinguish solid from liquid-like particles $\bar{q}_{6,t}=0.34$ and the 
dashed-dotted red line represents the threshold to discriminate between ice Ic and ice Ih $\bar{q}_{4,t}=0.41$.} 
\label{optimizacion-q6q4}
\end{figure}

As previously mentioned, in order to get an interface profile, $h(x_n)$, we first need to identify the 
molecules belonging to the ice phase. This is done by means of the $\bar{q}_{l}$ order parameter
proposed by Lechner and Dellago \cite{lechnerDellago08}. The order parameter is a scalar number
that is assigned to each molecule according to the degree of orientational order 
in its local environment. 
In Fig. \ref{optimizacion-q6q4} we plot
$\bar{q}_{6}$ versus $\bar{q}_{4}$ for 3600 bulk molecules of liquid water (magenta), of ice-Ih (cyan) and
of ice-Ic (orange). Clearly, $\bar{q}_{6}$ is a good parameter to distinguish the fluid from either ice polymorph. 
The $\bar{q}_{6}$ threshold we use for that purpose is $\bar{q}_{6,t} = 0.34$ (horizontal dashed line in Fig. \ref{optimizacion-q6q4}). Thus, if 
a particle has a $\bar{q}_{6}$ value larger than $\bar{q}_{6,t}$ it is labelled as solid-like, and vice-versa. 
As it can be seen in Fig. \ref{optimizacion-q6q4}, molecules belonging to ice-Ih and ice-Ic polymorphs
can be distinguished with the $\bar{q}_{4}$ order parameter with a threshold of $\bar{q}_{4,t} = 0.41$ (vertical dashed-dotted line
in Fig. \ref{optimizacion-q6q4}). 

\begin{figure}

\begin{subfigure}[b]{0.15\paperheight}
\resizebox{1.5in}{!}{\includegraphics{fig14.eps}}
\caption{}
\end{subfigure}
\begin{subfigure}[b]{0.15\paperheight}
\resizebox{1.5in}{!}{\includegraphics{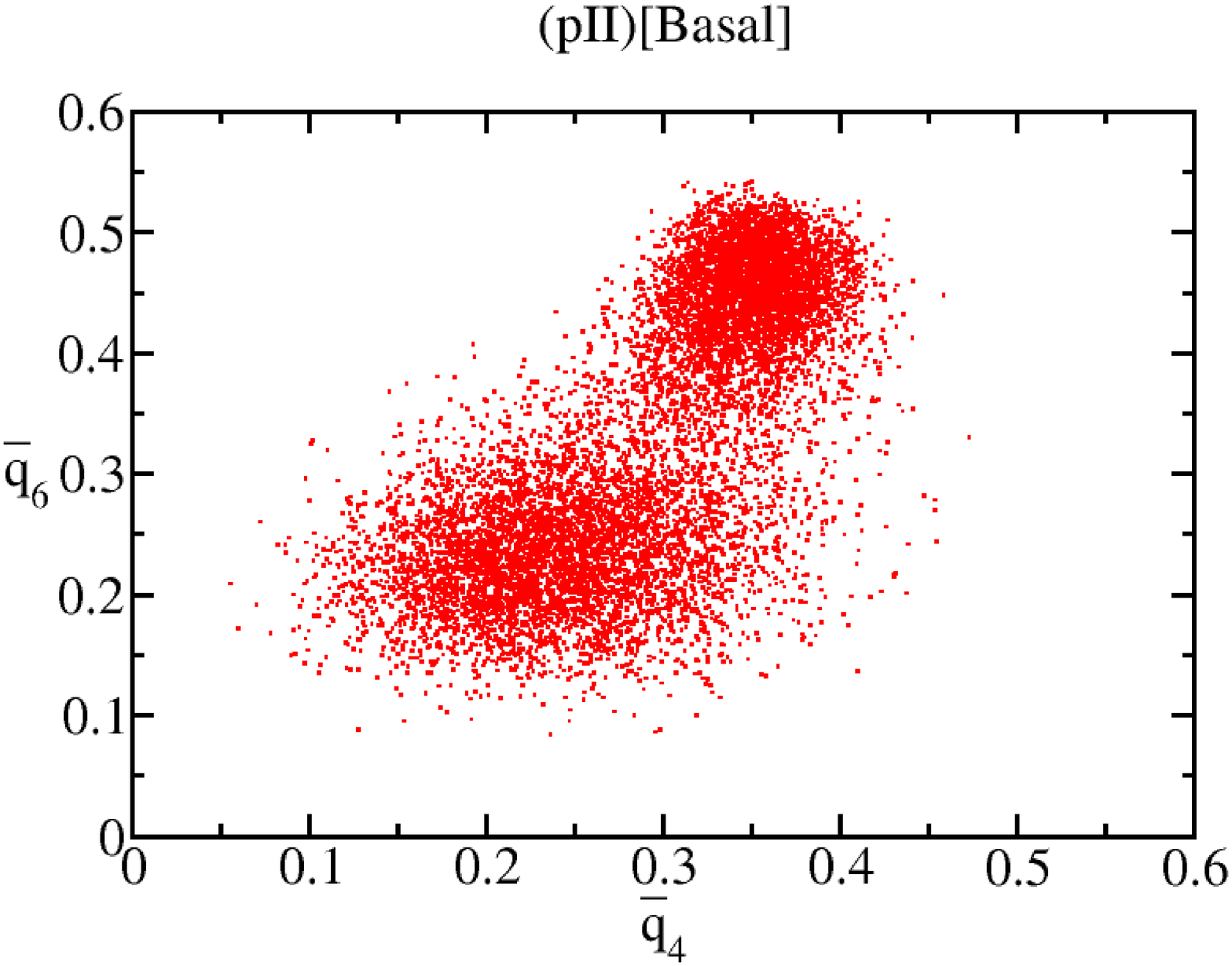}}
\caption{}
\end{subfigure}
\begin{subfigure}[b]{0.15\paperheight}
\resizebox{1.5in}{!}{\includegraphics{fig16.eps}}
\caption{}
\end{subfigure}
\begin{subfigure}[b]{0.15\paperheight}
\resizebox{1.5in}{!}{\includegraphics{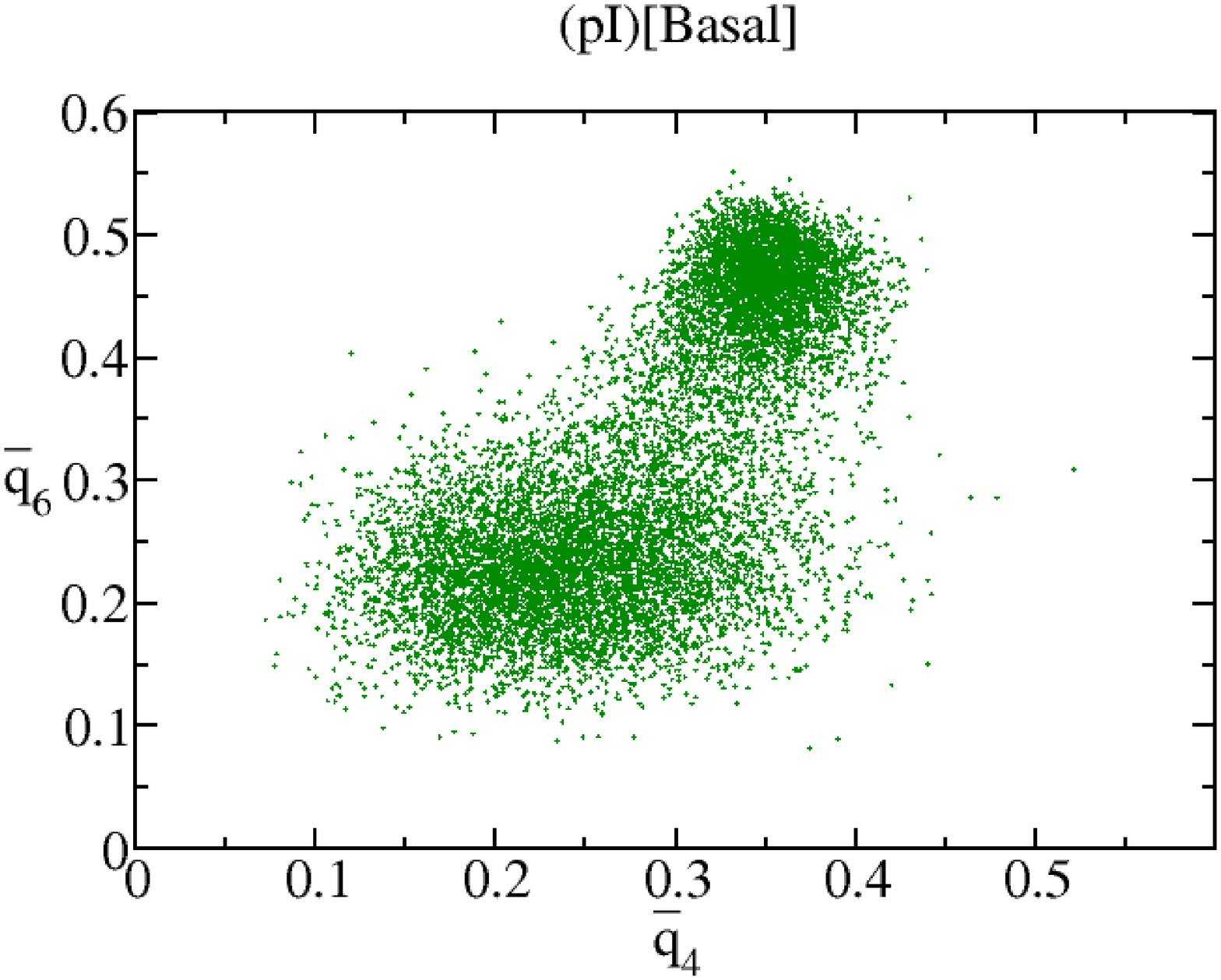}}
\caption{}
\end{subfigure}
\begin{subfigure}[b]{0.15\paperheight}
\resizebox{1.5in}{!}{\includegraphics{fig18.eps}}
\caption{}
\end{subfigure}
 \caption{$\bar q_{6}-\bar q_{4}$ maps for the 
 last configuration of each of the systems studied. When the basal plane
is exposed, panel (e), a cloud of points at high $\bar q_{4}$ corresponding
to ice-Ic emerges.}
 \label{q6q4}
\end{figure}

\begin{figure}
\includegraphics[width=0.40\paperwidth,height=0.33\paperwidth,keepaspectratio]{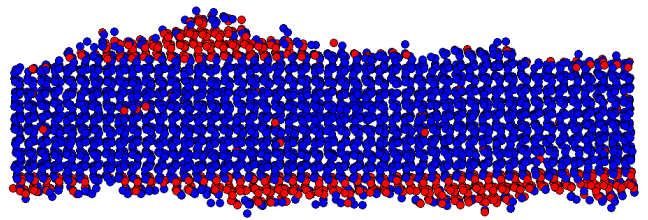}
\caption{Oxygens of the molecules in the ice slab for the system in which 
the basal plane is exposed to the liquid. Blue:
oxygen atoms with ice-Ih environment. Red: oxygen atoms with ice Ic 
environment.}
 \label{hielo-ic}
\end{figure}

Initially, the system is prepared by putting an ice-Ih slab in contact with liquid
water. Therefore, by analysing the $\bar{q}_{6}-\bar{q}_{4}$ map of the initial configuration
one would obtain points in the region of the pink and the cyan clouds of Fig. \ref{optimizacion-q6q4}.
At the end of the simulations all orientations where a prismatic plane is exposed only show 
these two clouds of points (see Fig. \ref{q6q4} a-d). Therefore, there is only liquid and ice Ih at 
the end of these simulations. However, the
simulation where the basal plane is exposed to the liquid shows an extra cloud of points in the area corresponding to 
ice-Ic (Fig. \ref{q6q4} e). This suggests that some molecules with ice-Ic environment
appear along the course of the simulation. To know where these molecules are  located we plot in Fig. \ref{hielo-ic} ice molecules with
$\bar{q}_{4,t} < 0.41$ in blue (ice Ih) and with $\bar{q}_{4,t} > 0.41$ in red (ice Ic). 
Clearly, thin ice-Ic layers have developed on some regions of the ice-water interface. 
In Ref. \cite{benetjcp2014} we show that the relaxation of crystal-fluid capillary waves is due 
to the continuous recrystallization and melting taking place at the interface. 
This relaxation mechanism allows for the epitaxial growth of ice Ic on top of the underlying 
ice Ih. 
The recrystallization/melting relaxation mechanism also explains our observation that
the interfacial regions containing ice Ih and ice Ic dynamically change along the course of the simulation. 
The reason why this structural transformation is only present when the basal 
plane is exposed is that hexagonal and cubic ice differ in their stacking sequence 
along the direction perpendicular to the basal plane (Ice Ic stacking 
is diamond-like, A,B,C,A,B,C,... whereas ice Ih is wurtzite-like, A,B,A,B,...). Therefore, when the basal plane is exposed
an Ic-stacking can grow on top of ice Ih, but the same is not true for the prismatic planes. 
By analysing a set of over 300 configurations with the basal plane 
exposed we observe that about  60 \% of the ice in contact with water is Ic and
the other 40 \% is ice Ih. This is not altogether unexpected, since, at least
for the TIP4P models, the free energy of ice Ic is
very similar to that of ice Ih \cite{PRL_2004_92_255701}. Accordingly, growth of regions of
ice Ic with a negligible bulk free energy penalty can be realized if the
corresponding surface free energy of the newly formed Ih-Ic and Ic-water interfaces
is comparable to that of the bare Ih-water interface. 
The phenomenon above described resembles 
preferential adsorption of a metastable phase, well known in a variety
of systems,\cite{cao90,mueller02} as well as in the ice-vapour interface, 
which is mediated by a thin water layer \cite{elbaum91,wilen95,libbrecht05}. 
However, in the case here studied ice-Ic does not fully cover the interface but 
dynamically coexists at the interface with ice-Ih.

Therefore, our simulations predict that both ice polymorphs live together
in the interface at equilibrium. This is not the only situation in which hexagonal and cubic ice 
can be found in close contact:
there is compelling experimental and simulation evidence that ice grows with a mixed Ic-Ih
stacking from supercooled water \cite{Malkinpnas2012,moorepccp2011,seojcp2012,carignanojpcc2007,kusalik2011} or from vapour deposition \cite{Kuhspnas2012}. 

\section{Conclusions and outlook}
In this work we use the TIP4P/2005 water model to study the ice-water interface by means of computer simulations. 
We simulate the ice-water interface under coexistence conditions and evaluate the interfacial stiffness and the interfacial free energy
by measuring the spectrum of capillary fluctuations. We study different crystal orientations and wave propagation 
directions. The predictions we get from the TIP4P/2005 model are the following:

\begin{itemize}
\item{The orientationally averaged interfacial free energy is 27(2) mN/m, in fair agreement with that obtained by analysing, for the same model, the
size of critical ice nuclei with Classical Nucleation Theory \cite{jacs2013}. Our value is also similar to that obtained for other TIP4P family models
by means of a cleaving methodology \cite{doi:10.1021/ct300193e} and is consistent with experimental estimates of the interfacial free energy that range 
from 25 to 35 mN/m \cite{pruppacher1995}.}
\item{We obtain an interfacial free energy of 27(2), 28(2), and 28(2) mN/m for 
the basal, primary prismatic and secondary prismatic 
planes respectively. The accuracy of our calculations is not enough to establish definite conclusions about the anisotropy of the interfacial free energy, but our 
results suggest, in accordance with predictions for other TIP4P family models \cite{doi:10.1021/ct300193e}, that the basal plane has the lowest free energy.}
\item{By measuring the density along the direction perpendicular to the
interface we estimate an upper bound for  the width of the ice-water interface
 of $\sim$ 4-5 molecular diameters, in fair agreement with the 3 molecular diameters obtained for the TIP4P model \cite{haymetjcp1988}.}
\item{
The ice-water interface for the basal plane shows alternating ice-Ih/ice-Ic regions. These change
dynamically due to capillary fluctuations.}
\end{itemize}

In a future, it would be useful to explore how to improve the 
accuracy of the present methodology in order to
capture the small anisotropy of the ice-water interfacial free energy. 
Moreover, the study of other water models could improve our understanding
on the ice-water interface. Of particular interest is perhaps the TIP4P/ICE
model \cite{JCP_2005_122_234511}, whose melting properties are close to those of real water. 
However, we do not expect large differences with the values reported
here for the TIP4P/2005 given the similarity between
all TIP4P family models studied so far (TIP4P, TIP4P-Ew and TIP4P/2005).
On the other hand, it would be interesting to pursue a more quantitative
analysis of the coexistence of cubic and hexagonal ice patches at the
ice-water interface, like, e.g., a characterization of the typical 
size and relaxation times of such regions.  

{\bf Acknowledgements}\\
E. Sanz and J. Benet acknowledge financial support from the EU grant 322326-COSAAC-FP7-PEOPLE-2012-CIG
and from a Spanish grant Ramon y Cajal. L.G. MacDowell and J. Benet also
acknowledge financial support from project FIS2010-22047-C05-05 (Ministerio de
Economia y Competitividad).

\clearpage


\begin{thebibliography}{10}

\bibitem{pruppacher1995}
H.~R. Pruppacher, ``A new look at homogeneous ice nucleation in supercooled
  water drops,'' {\em J. Atmosph. Sci.}, vol.~52, p.~1924, 1995.

\bibitem{pruppacher1967}
H.~R. Pruppacher, ``Interpretation of experimentally determined growth rates of
  ice crystals in supercooled water,'' {\em The Journal of Chemical Physics},
  vol.~47, no.~5, pp.~1807--1813, 1967.

\bibitem{expsurfacetensionwater}
H.~J. White, J.~V. Sengers, D.~B. Neumann, and J.~C. Bellows, {\em Release on
  the Surface Tension of Ordinary Water Substance}.
\newblock IAPWS, 1995.

\bibitem{doi:10.1021/ct300193e}
R.~L. Davidchack, R.~Handel, J.~Anwar, and A.~V. Brukhno, ``Ice ih-water
  interfacial free energy of simple water models with full electrostatic
  interactions,'' {\em Journal of Chemical Theory and Computation}, vol.~8,
  no.~7, pp.~2383--2390, 2012.

\bibitem{PhysRevLett.100.036104}
R.~Handel, R.~L. Davidchack, J.~Anwar, and A.~Brukhno, ``Direct calculation of
  solid-liquid interfacial free energy for molecular systems: Tip4p ice-water
  interface,'' {\em Phys. Rev. Lett.}, vol.~100, p.~036104, Jan 2008.

\bibitem{broughton:5759}
J.~Q. Broughton and G.~H. Gilmer, ``Molecular dynamics investigation of the
  crystal--fluid interface. vi. e xcess surface free energies of
  crystal--liquid systems,'' {\em J. Chem. Phys.}, vol.~84, no.~10,
  pp.~5759--5768, 1986.

\bibitem{JCP_2005_123_234505}
J.~L.~F. Abascal and C.~Vega, ``A general purpose model for the condensed
  phases of water: Tip4p/2005,'' {\em J. Chem. Phys.}, vol.~123, p.~234505,
  2005.

\bibitem{Vega11}
C.~Vega and J.~L.~F. Abascal, ``Simulating water with rigid non-polarizable
  models: a general perspective,'' {\em Phys. Chem. Chem. Phys.}, vol.~13,
  pp.~19663--19688, 2011.

\bibitem{jacs2013}
E.~Sanz, C.~Vega, J.~R. Espinosa, R.~Caballero-Bernal, J.~L.~F. Abascal, and
  C.~Valeriani, ``Homogeneous ice nucleation at moderate supercooling from
  molecular simulation,'' {\em Journal of the American Chemical Society},
  vol.~135, no.~40, pp.~15008--15017, 2013.

\bibitem{bai:124707}
X.-M. Bai and M.~Li, ``Calculation of solid-liquid interfacial free energy: A
  classical nucleation theory based approach,'' {\em J. Chem. Phys.}, vol.~124,
  no.~12, p.~124707, 2006.

\bibitem{ZPC_1926_119_277_nolotengo}
M.~Volmer and A.~Weber {\em Z. Phys. Chem.}, vol.~119, p.~277, 1926.

\bibitem{becker-doring}
R.~Becker and W.~Doring {\em Ann. Phys.}, vol.~24, pp.~719--752, 1935.

\bibitem{PhysRevLett.86.5530}
J.~J. Hoyt, M.~Asta, and A.~Karma, ``Method for computing the anisotropy of the
  solid-liquid interfacial free energy,'' {\em Phys. Rev. Lett.}, vol.~86,
  pp.~5530--5533, Jun 2001.

\bibitem{JCP_2006_125_094710}
R.~L. Davidchack, J.~R. Morris, and B.~B. Laird, ``The anisotropic hard-sphere
  crystal-melt interfacial free energy from fluctuations,'' {\em J. Chem.
  Phys.}, vol.~125, p.~094710, 2006.

\bibitem{morris:3920}
J.~R. Morris and X.~Song, ``The anisotropic free energy of the lennard-jones
  crystal-melt interface,'' {\em J. Chem. Phys.}, vol.~119, no.~7,
  pp.~3920--3925, 2003.

\bibitem{wangmorrisJCP2013}
J.~Wang, P.~A. Apte, J.~R. Morris, and X.~C. Zeng, ``Freezing point and
  solid-liquid interfacial free energy of stockmayer dipolar fluids: A
  molecular dynamics simulation study,'' {\em The Journal of Chemical Physics},
  vol.~139, no.~11, p.~114705, 2013.

\bibitem{lechnerDellago08}
W.~Lechner and C.~Dellago, ``Accurate determination of crystal structures based
  on averaged local bond order parameters,'' {\em The Journal of Chemical
  Physics}, vol.~129, no.~11, p.~114707, 2008.

\bibitem{benetjcp2014}
J.~Benet, L.~G. MacDowell, and E.~Sanz, ``Computer simulation study of surface
  wave dynamics at the crystal--melt interface,'' {\em J. Chem. Phys.},
  vol.~141, p.~024307, 2014.

\bibitem{fisher83}
D.~S. Fisher and J.~D. Weeks, ``Shape of crystals at low temperatures: Absence
  of quantum roughening,'' {\em Phys. Rev. Lett.}, vol.~50, pp.~1077--1080, Apr
  1983.

\bibitem{privman92}
V.~Privman, ``Fluctuating interfaces, surface tension and capillary waves: An
  introduction,'' {\em International Journal of Modern Physics C}, vol.~3,
  pp.~857--877, 1992.

\bibitem{jasnow84}
D.~Jasnow, ``Critical phenomena at interfaces,'' {\em Rep. Prog. Phys.},
  vol.~47, no.~9, p.~1059, 1984.

\bibitem{nelson04}
D.~Nelson, T.~Piran, and S.~Weinberg, {\em Statistical Mechanics of Membranes
  and Surfaces}.
\newblock Word Scientific, Singapore, 2004.

\bibitem{kara81}
M.~Kara and K.~Kurki-Suonio, ``Symmetrized multipole analysis of orientational
  distributions,'' {\em Acta Crystallographica Section A}, vol.~37, no.~2,
  pp.~201--210, 1981.

\bibitem{GROMACS1}
H.~Berendsen, D.~van~der Spoel, and R.~van Drunen, ``Gromacs: A message-passing
  parallel molecular dynamics implementation,'' {\em Computer Physics
  Communications}, vol.~91, no.~1–3, pp.~43 -- 56, 1995.

\bibitem{GROMACS2}
B.~Hess, C.~Kutzner, D.~van~der Spoel, and E.~Lindahl, ``Gromacs 4: Algorithms
  for highly efficient, load-balanced, and scalable molecular simulation,''
  {\em Journal of Chemical Theory and Computation}, vol.~4, no.~3,
  pp.~435--447, 2008.

\bibitem{bussi07}
G.~Bussi, D.~Donadio, and M.~Parrinello, ``Canonical sampling through velocity
  rescaling,'' {\em J. Chem. Phys.}, vol.~126, no.~1, p.~014101, 2007.

\bibitem{Frenkel_darkside}
D.~Frenkel, ``Simulations: the dark side,'' {\em Eur. Phys. J. Plus}, vol.~128,
  p.~10, 2013.

\bibitem{0295-5075-93-2-26006}
R.~E. Rozas and J.~Horbach, ``Capillary wave analysis of rough solid-liquid
  interfaces in nickel,'' {\em EPL (Europhysics Letters)}, vol.~93, no.~2,
  p.~26006, 2011.

\bibitem{maruyama97}
M.~Maruyama, T.~Nishida, and T.~Sawada, ``Crystal shape of high-pressure ice ih
  in water and roughening transition of the (101̄0) plane,'' {\em The Journal
  of Physical Chemistry B}, vol.~101, no.~32, pp.~6151--6153, 1997.

\bibitem{sun06}
D.~Y. Sun, M.~I. Mendelev, C.~A. Becker, K.~Kudin, T.~Haxhimali, M.~Asta, J.~J.
  Hoyt, A.~Karma, and D.~J. Srolovitz, ``Crystal-melt interfacial free energies
  in hcp metals: A molecular dynamics study of mg,'' {\em Phys. Rev. B},
  vol.~73, p.~024116, Jan 2006.

\bibitem{hardy77}
S.~C. Hardy, ``A grain boundary groove measurement of the surface tension
  between ice and water,'' {\em Philosophical Magazine}, vol.~35, no.~2,
  pp.~471--484, 1977.

\bibitem{weeks77}
J.~D. Weeks, ``Structure and thermodynamics of the liquid--vapor interface,''
  {\em J. Chem. Phys.}, vol.~67, no.~7, pp.~3106--3121, 1977.

\bibitem{ocko94}
B.~M. Ocko, X.~Z. Wu, E.~B. Sirota, S.~K. Sinha, and M.~Deutsch, ``X-ray
  reflectivity study of thermal capillary waves on liquid surfaces,'' {\em
  Phys. Rev. Lett.}, vol.~72, pp.~242--245, Jan 1994.

\bibitem{werner97}
A.~Werner, F.~Schmid, M.~Muller, and K.~Binder, ``Anomalous size-dependence of
  interfacial profiles between coexisting ph ases of polymer mixtures in
  thin-film geometry: A monte carlo simulation,'' {\em J. Chem. Phys.},
  vol.~107, no.~19, pp.~8175--8188, 1997.

\bibitem{macdowell14}
L.~G. MacDowell, J.~Benet, N.~A. Katcho, and J.~M. Palanco, ``Disjoining
  pressure and the film-height-dependent surface tension of thin liquid films:
  New insight from capillary wave fluctuations,'' {\em Advances in Colloid and
  Interface Science}, vol.~206, no.~0, pp.~150--171, 2014.

\bibitem{haymetjcp1988}
O.~A. Karim and A.~D.~J. Haymet, ``The ice/water interface: A molecular
  dynamics simulation study,'' {\em The Journal of Chemical Physics}, vol.~89,
  no.~11, pp.~6889--6896, 1988.

\bibitem{PRL_2004_92_255701}
E.~Sanz, C.~Vega, J.~L.~F. Abascal, and L.~G. MacDowell, ``Phase diagram of
  water from computer simulation,'' {\em Phys. Rev. Lett.}, vol.~92, p.~255701,
  2004.

\bibitem{cao90}
Y.~Cao and E.~H. Conrad, ``Approach to thermal roughening of ni(110): A study
  by high-resolution low-energy electron diffraction,'' {\em Phys. Rev. Lett.},
  vol.~64, pp.~447--450, Jan 1990.

\bibitem{mueller02}
M.~M{\"u}ller, L.~G. MacDowell, P.~Virnau, and K.~Binder, ``Interface
  properties and bubble nucleation in compressible mixtures containing
  polymers,'' {\em J. Chem. Phys.}, vol.~117, pp.~5480--5496, 2002.

\bibitem{elbaum91}
M.~Elbaum, ``Roughening transition observed on the prism facet of ice,'' {\em
  Phys. Rev. Lett.}, vol.~67, pp.~2982--2985, Nov 1991.

\bibitem{wilen95}
L.~A. Wilen, J.~S. Wettlaufer, M.~Elbaum, and M.~Schick, ``Dispersion-force
  effects in interfacial premelting of ice,'' {\em Phys. Rev. B}, vol.~52,
  pp.~12426--12433, Oct 1995.

\bibitem{libbrecht05}
K.~G. Libbrecht, ``The physics of snow crystals,'' {\em Rep. Prog. Phys.},
  vol.~68, pp.~855--895, 2005.

\bibitem{Malkinpnas2012}
T.~L. Malkin, B.~J. Murray, A.~V. Brukhno, J.~Anwar, and C.~G. Salzmann,
  ``Structure of ice crystallized from supercooled water,'' {\em Proceedings of
  the National Academy of Sciences}, vol.~109, no.~4, pp.~1041--1045, 2012.

\bibitem{moorepccp2011}
E.~B. Moore and V.~Molinero, ``Is it cubic? ice crystallization from deeply
  supercooled water,'' {\em Phys. Chem. Chem. Phys.}, vol.~13,
  pp.~20008--20016, 2011.

\bibitem{seojcp2012}
M.~Seo, E.~Jang, K.~Kim, S.~Choi, and J.~S. Kim, ``Understanding anisotropic
  growth behavior of hexagonal ice on a molecular scale: A molecular dynamics
  simulation study,'' {\em The Journal of Chemical Physics}, vol.~137, no.~15,
  p.~154503, 2012.

\bibitem{carignanojpcc2007}
M.~A. Carignano, ``Formation of stacking faults during ice growth on hexagonal
  and cubic substrates,'' {\em The Journal of Physical Chemistry C}, vol.~111,
  no.~2, pp.~501--504, 2007.

\bibitem{kusalik2011}
D.~Rozmanov and P.~G. Kusalik, ``Temperature dependence of crystal growth of
  hexagonal ice (ih),'' {\em Phys. Chem. Chem. Phys.}, vol.~13,
  pp.~15501--15511, 2011.

\bibitem{Kuhspnas2012}
W.~F. Kuhs, C.~Sippel, A.~Falenty, and T.~C. Hansen, ``Extent and relevance of
  stacking disorder in “ice ic”,'' {\em Proceedings of the National Academy
  of Sciences}, vol.~109, no.~52, pp.~21259--21264, 2012.

\bibitem{JCP_2005_122_234511}
J.~L.~F. Abascal, E.~Sanz, R.~G. Fernandez, and C.~Vega, ``A potential model
  for the study of ices and amorphous water: {{TIP4P}/Ice},'' {\em J. Chem.
  Phys.}, vol.~122, p.~234511, 2005.

\end{thebibliography}

\end{document}